\documentclass[11pt,a4paper]{article}

\usepackage{bold-extra}

\usepackage{graphicx}

\graphicspath{ {./Figures/} }

\usepackage{comment}

\pdfoutput=1
\usepackage{color}
\usepackage{todonotes}
\usepackage{subfigure}
\usepackage{booktabs}
\usepackage{array}
\usepackage{multirow,rotating}
\usepackage{amsmath,amssymb,bm,graphicx,wasysym,mathrsfs,dsfont}
\usepackage{cite}
\usepackage{slashed,relsize}
\usepackage[colorlinks=true,linkcolor=black,anchorcolor=black,citecolor=blue,filecolor=cyan,menucolor=red,runcolor=filecolor,urlcolor=blue,bookmarks=true,bookmarksnumbered=true]{hyperref}

\usepackage[font=small,labelfont=bf]{caption}

\def\beq{\begin{equation}\displaystyle\displaystyle}
	\def\eeq{\end{equation}}
\def\bea{\begin{eqnarray}\displaystyle} 
	\def\eea{\end{eqnarray}}

\def\({\left(}
\def\){\right)}
\def\bry{\begin{array}}
	\def\ery{\end{array}}

\def\f{\frac}

\title{Collision Integrals for Cosmological Phase Transitions 
	\vspace{1cm}}
\date{}
\author{
	{\large Stefania De Curtis$^a$, Luigi Delle Rose$^b$, Andrea Guiggiani$^a$, \'Angel Gil Muyor$^c$,}\protect\\
	{\large and Giuliano Panico$^a$}\\
	[7mm]
	\normalsize\itshape $^a$ INFN Sezione di Firenze and Dipartimento di Fisica e Astronomia, \protect\\ Universit\`a di Firenze, Via G. Sansone 1, I-50019 Sesto Fiorentino, Italy\\
	\normalsize\itshape $^b$ Dipartimento di Fisica, \protect Universit\`a della Calabria, I-8703 Arcavacata di Rende, Cosenza, Italy\\
	\normalsize\itshape $^c$ IFAE and BIST, Universitat Aut\`onoma de Barcelona, 08193~Bellaterra,~Barcelona,~Spain
}

\begin{document}
	\baselineskip=14pt
	\arraycolsep=2pt
	
	\begin{flushright}
		$ $
	\end{flushright}
	
	\vspace{2em}
	
	{\let\newpage\relax\maketitle}
	\begin{abstract}
		\medskip
		\noindent
The dynamics of the true-vacuum bubbles nucleated during a first-order phase transition is affected by the distribution functions of the particle species in the plasma, driven out-of-equilibrium by the travelling domain wall.
An accurate modelling of this phenomenon is relevant for a quantitative description of phase transitions in the early universe and for the determination of the corresponding cosmic relics, such as, among the others, the stochastic background of gravitational waves.
We address this problem by developing a new spectral method devised for a fast and reliable computation of the collision integral in the Boltzmann equations.
In a scalar singlet extension of the Standard Model chosen as a benchmark scenario, we test our algorithm, determining the bubble speed and profile, and we asses the impact of the out-of-equilibrium dynamics.  

	\end{abstract}
	
	\vfill
	\noindent\line(1,0){188}\\
	{\scriptsize{E-mail: \texttt{stefania.decurtis@fi.infn.it}, \texttt{luigi.dellerose@unical.it}, \texttt{andrea.guiggiani@unifi.it},\\ \texttt{agil@ifae.es}, \texttt{giuliano.panico@unifi.it}}}
	
	\thispagestyle{empty}
	
	\newpage
	
	\begingroup
	\tableofcontents
	\endgroup 
	
	\setcounter{equation}{0}
	\setcounter{footnote}{0}
	\setcounter{page}{1}
	
	\newpage
	
	\section{Introduction}\label{sec:intro}

First order phase transitions (PhTs) in the early Universe  can lead to striking cosmological signatures including, for example, a stochastic background of gravitational waves or a significant departure from equilibrium required by scenarios of electroweak (EW) baryogenesis to generate the observed matter-antimatter asymmetry. Future gravitational wave interferometers, such as LISA~\cite{Caprini:2015zlo,Caprini:2019egz}, DECIGO~\cite{Kawamura:2006up,Kawamura:2011zz}, Taiji~\cite{Hu:2017mde,Ruan:2018tsw} and TianQuin~\cite{TianQin:2015yph}, have the potential to detect the aforementioned background and thus to shed light on the nature of the EWPhT, providing a new boost in the quest for the physics beyond the Standard Model (BSM).

In order to accurately describe the power spectrum of the gravitational waves and to reconstruct from that the properties of the early Universe at the time of the transition, we need a precise description of the bubble dynamics. Indeed, during a first-order PhT, bubbles of the stable phase nucleate and expand through space interacting with the surrounding plasma.
One of the most relevant parameters describing this dynamics is the speed of the domain wall (DW) in the steady state regime. This affects both the magnitude and the shape of the gravitational wave spectrum and, most probably, it will be the parameter determined with the best accuracy by the future generation of interferometers~\cite{Gowling:2021gcy}. 

The first computation of the bubble speed dates back to refs.~\cite{Moore:1995ua,Moore:1995si}. In those works the friction exerted on the DW by the particles of the plasma was computed from their out-of-equilibrium distribution functions, which are obtained, in turn, as solutions to the corresponding Boltzmann equations. These results were fed to the equation of motion of the DW which was then used to extract the bubble speed and the bubble width.
 
The challenging part of the computation resides in the Boltzmann equation and, in particular, in the collision integral that describes the interactions among the particles in the plasma.
This term represents the bottleneck of these kind of computations since it makes the Boltzmann equation a complicated integro-differential problem.  
The strategy exploited in ref.~\cite{Moore:1995si} to address this issue was based on the fluid approximation which corresponds to a first order momentum expansion of the out-of-equilibrium distributions. Within this approach, these are parametrized by three space-dependent perturbations: the chemical potential, the temperature and the velocity fluctuations.
By taking moments of the Boltzmann equation with suitably chosen weight factors, the fluid approximation allows one to turn the integro-differential equation into a simpler system of ordinary differential equations.
	
Since this formalism strongly relies on the choice of the weight basis and on a specific ansatz for the momentum dependence of the distribution functions, it is intrinsically affected by some degree of arbitrariness. Moreover, a first order truncation may not be sufficient to capture all the relevant features of the out-of-equilibrium distributions. Indeed, subsequent works~\cite{Laurent:2020gpg,Dorsch:2021ubz,DeCurtis:2022hlx} have shown that the fluid approximation is not particularly reliable, neither quantitatively nor qualitatively. 
	
With the successes of precision cosmology and the new opportunities offered by the gravitational wave interferometry, the necessity of a solid description of the bubble dynamics in a first order PhT is more than pressing. In this respect, a huge effort has been done recently to conceive more reliable approximation methods and to develop a deeper theoretical understanding. Several attempts have been carried out to ameliorate the existing results. For instance, the fluid approximation has been extended in refs.~\cite{Dorsch:2021ubz,Dorsch:2021nje} with the inclusion of higher orders in the small momentum expansion; a less constrained momentum dependence has been exploited in refs.~\cite{Laurent:2020gpg,Cline:2000nw} by using a factorization ansatz. 
In ref.~\cite{Laurent:2022jrs} an improved and faster algorithm has been proposed which exploits a decomposition of the out-of-equilibrium distributions on a basis of Chebyshev polynomials. \\ 
For the sake of completeness, we also mention that other approaches, based on phenomenological modelling of the friction through a viscosity parameter, have also been explored~\cite{Megevand:2009gh,Espinosa:2010hh,Leitao:2010yw,Megevand:2013hwa,Huber:2013kj,Megevand:2013yua,Leitao:2014pda,Megevand:2014yua,Megevand:2014dua}.
	
In ref.~\cite{DeCurtis:2022hlx,DeCurtis:2022djw,DeCurtis:2022llw} we made a step forward in the development of an accurate method to determine the out-of-equilibrium distributions without imposing any specific momentum dependence. We presented, for the first time, a fully quantitative solution to the Boltzmann equation designed to compute the collision integrals through an iterative algorithm without the use of any ansatz. 
This methodology can be applied to evaluate the friction on the bubble wall and the terminal speed of the latter, but it can be used also to describe the non-equilibrium properties of the distribution functions that are relevant for the determination of the matter-antimatter asymmetry in models of baryogenesis. 
The algorithm discussed in ref.~\cite{DeCurtis:2022hlx} allows one to compute the deviations from equilibrium of the particles in the plasma, as function of their momenta, induced by the presence of a travelling DW. The system is mainly characterized by the bubble speed $v_w$, the wall width $L_w$ and the non-zero vacuum expectation value $v$ of the scalar driving the PhT. These parameters are taken as input of the iterative procedure.

While the proposed algorithm for the solution of the Boltzmann equation can reach convergence within a small number of steps, it can quickly become computationally expensive when it is embedded in the full algorithm that scans over the space of $v_w$, $L_w$ and $v$. This algorithm is used to identify the values of the parameters that simultaneously solve the Boltzmann equation and the equation of motion of the DW. 
The procedure would be even more time-demanding if one explores the parameter space of new physics models in the search for optimal points providing observable signals of gravitational waves as well as, for instance, the correct amount of the baryon asymmetry.

Analogously to previous works, the bottleneck of our iterative algorithm is the computation of the collision integral. Although some algebraic manipulations were used to reduce the dimensionality of the integral, the large number of integrations, together with the singular behavior of the integration kernels, strongly limited the computational speed, even on a cluster.
In the present work we significantly improve on our previous results by providing a more efficient method for the computation of the collision integral. 
We exploit a spectral decomposition of the collision operator in terms of its eigenfunctions, effectively reducing a complex and time-consuming nine-dimensional integration in a much faster matrix multiplication. Moreover, the eigenfunctions are computed only once and can be reused during all the scanning procedure of an entire model.
	
For the purpose of presenting our new methodology, we consider the case of a first order EWPhT driven by two scalar fields. This scenario can be realized by enlarging  the SM particle spectrum with a singlet scalar field coupled to the SM sector only through a quartic portal interaction. For the computation of the friction we focus on the contribution of the top quark species which has the largest coupling, among all the SM particles, to the Higgs field. All the other particles are assumed to be in local equilibrium and considered as background. The inclusion of the out-of-equilibrium distributions of other species in the computation of the friction, such as the massive EW gauge bosons, is straightforward and it will be addressed in a future work. 
	
The paper is organized as follows. In section~\ref{sec:boltzmann} we review the Boltzmann equation and we introduce the necessary notation while in section~\ref{sec:decomposition} we present the new spectral decomposition method. In section~\ref{sec:hydro} we discuss the hydrodynamic equations for the temperature and velocity profiles and in section~\ref{sec:analysis} we discuss the numerical results. The conclusions are drawn in section~\ref{sec:concl}. Appendix~\ref{sec:appendix} summarizes some technical details on the solutions to the hydrodynamic equations.

	\section{The Boltzmann equation}\label{sec:boltzmann}
 
    Our main goal is to compute the terminal velocity of the DW that surrounds the true-vacuum bubbles. This parameter is the result of a balance of the internal pressure, proportional to the potential energy difference between the two vacua, and the friction provided by particles in the plasma impinging on the DW. Usually, for the computation of the terminal velocity
    the relevant configurations are those in which the radius of the wall is much larger than its thickness. In such a case, as we will do in the following, we can adopt the planar limit, orient the $z$-axis along the propagating direction of the DW and assume that a steady state is reached. 

    The friction acting on the bubble wall is determined
    by the deviations of the distribution function from equilibrium, $\delta f$, and in the steady state and planar limit it is given by
    \begin{equation}\label{eq:friction}
        F(z) = \sum_i\f{N_i}{2}\f{dm^2_i}{dz}\int \f{d^3{\bf p}}{(2\pi)^3E_p}\delta f_i\,,
    \end{equation}
    where the sum runs over all the particle species in the plasma, $m_i$ is the mass of the $i$-th particle, $N_i$ its degrees of freedom and $\delta f_i$ the corresponding perturbation.

		The perturbations around equilibrium are computed by solving the Boltzmann equation for the distribution function of the plasma in the presence of an expanding bubble of true vacuum. 
		It is convenient to write the Boltzmann equation in the wall reference frame, where the solution is stationary.
        The equation for the distribution function $f$ of a particle species in the plasma is
		
		\begin{equation}\label{eq:boltz_eq}
			{\cal L}[f]\equiv\left(\frac{p_z}{E}\partial_z - \frac{(m^2(z))'}{2E}\partial_{p_z}\right) f = -{\cal C}[f]\,,
		\end{equation}
		where ${\cal L}$ is the Liouville operator, ${\cal C}$ is the collision operator, $m(z)$ is the mass of the particle and its derivative is performed along the $z$ direction.
        As we mentioned before, in this work we consider the top quark to be the only species out-of-equilibrium, since it is the one with the strongest interaction with the Higgs field.
        We assume that the light degrees of freedom of the plasma, namely all the SM degrees of freedom but the top, are in local equilibrium and we treat them as background fluids described by the standard Fermi-Dirac or Bose-Einstein distributions.
        
        For a perfect fluid, in the wall reference frame the distribution function is
		
		\begin{equation}
			f_v = \frac{1}{e^{\beta(z)\gamma_p(z)(E-v_p(z)p_z)}\pm 1}\,,
		\end{equation}
		where $\beta = T^{-1}$, $\gamma_p$ is the Lorentz gamma factor,
        and $v_p(z)$ is the velocity profile of the plasma measured in the wall reference frame. In local equilibrium, macroscopic quantities in general depend on the position $z$. 
  
  Since we consider small perturbations around equilibrium, we can linearize the Boltzmann equation in $\delta f$, namely $f = f_v + \delta f$. As we showed in ref.~\cite{DeCurtis:2022hlx} the linearized Boltzmann equation can be cast in the general form
		
		\begin{equation}\label{eq:boltz_S}
			{\cal L}[\delta f] - \frac{\cal Q}{E} \, \delta f = \frac{p_z}{E}{\cal S}\, + \langle \delta f\rangle\,,
		\end{equation}
        where the source term ${\cal S}$ originates from the action of the Liouville operator on $f_v$, while the bracket $\langle \delta f\rangle$ and ${\cal Q}$ arise from the linearized collision operator.
		
By considering $2\leftrightarrow 2$ scattering processes, the collision operator takes the following form
		
			\begin{equation}
				{\cal C}[f] = \sum_i \frac{1}{4 N_p E}\int \frac{d^3{\bf k}d^3{\bf p}'d^3{\bf k}'}{(2\pi)^5 2E_k 2E_{p'} 2E_{k'}}|{\cal M}_i|^2\delta^4(p+k-p'-k'){\cal P}[f]\,,
    \label{coll}
			\end{equation}
			with the sum performed over all the relevant processes with amplitude ${\cal M}_i$ and the population factor is given by
			\begin{equation}
				{\cal P}[f] = f(p)f(k)(1\pm f(p'))(1\pm f(k')) - f(p')f(k')(1\pm f(p))(1\pm f(k))\,.
			\end{equation}   
   In eq.~(\ref{coll}), $N_p$ represents the degrees of freedom of the incoming particle with momentum $p$, $k$ is the momentum of the second incoming particle, while $p'$ and $k'$ are the momenta of the particles in the final state. The $+$ sign is for bosons, while the $-$ sign is for fermions. After the linearization  in $\delta f$ the collision operator is given by
			\begin{equation}\label{eq:coll_lin}
				\bar{\cal C}[f] \equiv -\frac{{\cal Q}}{E}\frac{f_v}{f'_v}\delta f(p) - \langle \delta f\rangle=\sum_i \frac{1}{4 N_p E}\int \frac{d^3{\bf k}d^3{\bf p}'d^3{\bf k}'}{(2\pi)^5 2E_k 2E_{p'} 2E_{k'}}|{\cal M}_i|^2\delta^4(p+k-p'-k')\bar{\cal P}[f]\,,
			\end{equation}
			with 
			\begin{equation}
				\bar{\cal P}[f] = f_v(p)f_v(k)(1\pm f_v(p'))(1\pm f_v(k'))\sum\mp\frac{\delta f}{f_v'}\,,
			\end{equation}
			where the $\mp$ in the sum is for incoming and outgoing particles, respectively.
			As done in ref.~\cite{DeCurtis:2022hlx}, we can distinguish two contributions coming from the linearized collision operator: the first one, which depends only on $\delta f(p)$ so that the perturbation factorizes out of the integral, is described by the term ${\cal Q}/E \, \delta f$ in eq.~(\ref{eq:boltz_S}); the second one, denoted by the bracket $\langle \delta f\rangle$, includes the contributions where $\delta f$ appears under the integral sign.

	\section{Spectral decomposition of the collision operator}\label{sec:decomposition}

           The most computationally cumbersome term in the Boltzmann equation is the bracket $\langle \delta f\rangle$, which corresponds to a nine dimensional integral involving the unknown $\delta f$.
            One possible strategy to deal with this term is to perform, first, all the integrations that do not involve the unknown perturbation $\delta f$, as explained in ref.~\cite{DeGroot:1980dk}. The main advantage of this method is that such integrals depend only on the processes under consideration and have to be computed only once.
            
            Applying such procedure and assuming the particles to be massless inside the collision integral, the bracket term takes the following form~\cite{DeCurtis:2022hlx}
            \begin{equation}\label{eq:prel_bracket}
            \begin{split}
                \langle \delta f\rangle = -\frac{f_v(p)}{E_p}\bigg(\int\frac{d^3\bar{\bf k}}{2|\bar{\bf k}|} & \bigg[ f_0(\beta(z)|\bar{\bf k}|) \, {\cal K}_a(\beta(z)|\bar{\bf p}|, \beta(z)|\bar{\bf k}|,\theta_{\bar p \bar k}) \\
                &-  (1-f_0(\beta(z)|\bar{\bf k}|)) \, {\cal K}_s(\beta(z)|\bar{\bf p}|, \beta(z)|\bar{\bf k}|,\theta_{\bar p \bar k})\bigg]\frac{\delta f(k_\bot, k_z,z)}{f'_0(\beta(z)|\bar{\bf k}|)}\bigg)\,,
            \end{split}
            \end{equation}
            where barred momenta are computed in the local plasma reference frame, $f_0$ denotes the standard (Fermi--Dirac or Bose--Einstein) distribution functions, and
            $\theta_{\bar p \bar k}$ is the relative angle in the local plasma reference frame between the particles with momentum $p$ and $k$. The momenta $k_\bot$ and $k_z$, defined in the wall reference frame, can be expressed as functions of the local plasma momenta through a boost along the $z$-axis, namely $k_z = \gamma_p(z)(E_{\bar k} + v_p(z){\bar k}_z)$, $k_\bot = {\bar k}_\bot$. 
            The functions ${\cal K}_a$ and ${\cal K}_s$ are the annihilation and scattering kernels (we refer to ref.~\cite{DeCurtis:2022hlx} for the details)
            \begin{equation}\label{eq:kernel_eq}
            \begin{split}
                {\cal K}_a &= \frac{1}{8N_p(2\pi)^5}\int\frac{d^3{\bf k}'d^3{\bf p}'}{2E_{p'}2E_{k'}}|{\cal M}_a|^2(1\pm f_v(p'))(1\pm f_v(k'))\delta^4(p+k-p'-k')\\
                {\cal K}_s &= \frac{1}{8N_p(2\pi)^5}\int\frac{d^3{\bf k}d^3{\bf k}'}{2E_{k}2E_{k'}}|{\cal M}_s|^2f_v(k)(1\pm f_v(k'))\delta^4(p+k-p'-k') \,
            \end{split}
		      \end{equation}
            where the matrix elements ${\cal M}_a$ and ${\cal M}_s$ can be found in ref.~\cite{DeCurtis:2022hlx}.
            
            For a planar wall, the perturbation $\delta f$ depends only on the components of the momentum, $k_z$ and $k_\bot$,
             parallel and perpendicular to the propagating direction of the DW. As a consequence, eq.~(\ref{eq:prel_bracket}) can be further simplified exploiting the fact that the perturbation $\delta f$ is invariant under rotations around the $z$ axis. Using spherical coordinates $\{|\bar{\bf k}|, \theta_{\bar k}, \phi_{\bar k}\}$, where $\theta_{\bar k}$ is the polar angle between $\bar{\bf k}$ and the $z$ direction computed in the local plasma reference frame, it is possible to perform the integration over the azimuthal angle $\phi_{\bar k}$ because $\delta f$ does not depend on it. By also using the property
		      \begin{equation}
                (1\pm f_0(p)) = e^p f_0(p)\,,
		      \end{equation}
            we obtain
            \begin{equation}
    			\langle \delta f\rangle = -\frac{f_v(p)}{E_p}\int |\bar{\bf k}|d|\bar{\bf k}|d\cos \theta_{\bar k} f_0(\beta(z)|\bar{\bf k}|)\widetilde{\cal K}(\beta(z)|\bar{\bf p}|,\cos \theta_{\bar p},\beta(z)|\bar{\bf k}|,\cos \theta_{\bar k})\frac{\delta f(k_\bot, k_z, z)}{f_0'(\beta(z)|\bar{\bf k}|)}\,,
		      \end{equation}
            where $\widetilde {\cal K}$ is the result of the integration of the kernels over the azimuthal angle.
            
            Since the $z$ dependence in the kernel comes only from the temperature profile, i.e.~from the $\beta(z)$ factors, we can obtain a position independent expression through the local change of variables
            \begin{equation}
                \beta(z) k\rightarrow k \;\;\;\;\;\;\;\;\;\; \beta(z) p\rightarrow p \,.
            \end{equation}
            In this way, the $z$ dependence inside the integral is  encapsulated only in the perturbation $\delta f$ and we finally arrive to
            \begin{eqnarray}\label{eq:bracket_eq}
    			\langle \delta f\rangle &=& -\frac{f_v(p/\beta(z))}{\beta(z) E_p}\int |\bar{\bf k}|d|\bar{\bf k}|d\cos \theta_{\bar k} f_0(|\bar{\bf k}|)\widetilde{\cal K}(|\bar{\bf p}|,\cos \theta_{\bar p},|\bar{\bf k}|,\cos \theta_{\bar k})\frac{\delta f(k_\bot/\beta(z), k_z/\beta(z), z)}{f_0'(|\bar{\bf k}|)} \nonumber\\
                &\equiv& -\frac{f_v(p/\beta(z))}{\beta(z) E_p}\int {\cal D} \bar{k}\; \widetilde{\cal K}_{\bar{p},\bar{k}} \frac{\delta f(k_\bot/\beta(z), k_z/\beta(z), z)}{f_0'(|\bar{\bf k}|)} \,,
		      \end{eqnarray}
            where we introduced the notation ${\cal D} \bar{k} \equiv f_0(|\bar{\bf k}|) |\bar{\bf k}|d|\bar{\bf k}|d\cos \theta_{\bar k}$ and $\widetilde{\cal K}_{\bar{p},\bar{k}} \equiv \widetilde{\cal K}(|\bar{\bf p}|,\cos \theta_{\bar p},|\bar{\bf k}|,\cos \theta_{\bar k})$.

            The $\langle \delta f \rangle$ term  can be reinterpreted as the action of a Hermitian operator on the perturbation. For this purpose we define the operator
            \begin{equation}
                {\cal O}[g] \equiv \int {\cal D} \bar{k}\; \widetilde{\cal K}_{\bar{p},\bar{k}} \, g(|\bar{\bf k}|,\cos \theta_{\bar k})\,.
            \end{equation}
            Due to particle exchange symmetry, the kernel function $\widetilde{\cal K}$ is symmetric under the exchange $p \leftrightarrow k$, namely $\widetilde{\cal K}_{\bar{p},\bar{k}} = \widetilde{\cal K}_{\bar{k},\bar{p}}$. As a consequence, the operator ${\cal O}$ is Hermitian with respect to the scalar product
            \begin{equation}\label{eq:scalar_product}
            (f, g) = \int {\cal D} \bar{k}\, f(|\bar{\bf k}|,\cos \theta_{\bar k})\, g(|\bar{\bf k}|,\cos \theta_{\bar k})\,.
            \end{equation}
            This conclusion is clearly also valid for other choices of the scalar product. The one we use, as we will see in the following, is motivated by the fact that it significantly simplifies the evaluation of the collision integral.

            Thanks to its Hermiticity, the operator ${\cal O}$ can be diagonalized with an orthonormal basis of eigenfunctions
            \begin{equation}
            {\cal O}[\psi_l] = \int {\cal D}\bar{k}\, \widetilde{\cal K}_{\bar{p},\bar{k}}\, \psi_l(|\bar{\bf k}|, \cos \theta_{\bar{k}}) = \lambda_l\, \psi_l(|\bar{\bf k}|, \cos \theta_{\bar{k}})\,,
            \end{equation}
            and the kernel $\widetilde{\cal K}$ can be rewritten as \cite{1953mtp..book.....M}
            \begin{equation}
            \widetilde{\cal K}_{\bar{p},\bar{k}} = \sum_l \lambda_l\, \psi_l(|\bar{\bf p}|, \cos \theta_{\bar{p}})\, \psi_l(|\bar{\bf k}|, \cos \theta_{\bar{k}})\,.
            \end{equation}

		  Using this decomposition we can drastically simplify the computation of the bracket $\langle \delta f\rangle$. Indeed, once the eigenvectors are determined, the remaining two integrals involved in the computation become trivial due to the orthogonality property. The final expression for the bracket is then
    
		\begin{equation}\label{eq:bracket_decomposition}
			\langle \delta f\rangle = -\frac{f_v(p/\beta(z))}{\beta(z)E_p}\sum_l \lambda_l\, \phi_l(z)\, \psi_l(|\bar{\bf p}|, \cos \theta_{\bar p}),
		\end{equation}
		where $\phi_l(z)$ is the projection of the perturbation on the eigenstate basis, namely
        \begin{equation}\label{eq:perturbation_decomposition}
            \phi_l(z) =  \int {\cal D}\bar{k}\, \psi_l(|\bar{\bf k}|,\cos \theta_{\bar k})\frac{\delta f(k_\bot/\beta(z), k_z/\beta(z), z)}{f_0'(|\bar{\bf k}|)} \,.
		\end{equation}

As we will see in the next section, the main advantage of this decomposition method is the huge improvement in the timing performances in the computation of the bracket term.

\subsection{Numerical implementation}\label{subsec:numdec}

To implement numerically the spectral decomposition of the kernel, we need to choose a suitable basis of functions on which the perturbations can be expanded. A simple choice, which we use for our numerical analysis, is to discretize the $\{|\bar{\bf p}|, \cos\theta_{\bar p}\}$ space on a regular finite lattice. The functional space is then obtained from the discretized version through a suitable interpolation.

Using a rectangular lattice with $M$ and $N$ points in the $|\bar{\bf p}|$ and $\cos\theta_{\bar p}$ directions respectively, the operator ${\cal O}$ is represented by an $(MN) \times (MN)$ Hermitian matrix ${\cal U}$, which can be diagonalized to obtain the spectral decomposition. We computed the matrix elements of ${\cal U}$ on an orthonormal basis of functions $\{e_{mn}\}$ that vanish everywhere on the grid but on the point $(|\bar{\bf p}|_m,\cos\theta_{{\bar p}_n})$.

Some subtleties must be taken into account in the discretization process. The measure we adopted for the scalar product, ${\cal D} \bar{k}$, contains a factor $|\bar{\bf k}|$, which vanishes for $|\bar{\bf k}| = 0$. This means that if a zeroth-order (i.e.~a piecewise constant) approximation of the basis functions and of the integration measure is used, the elements $e_{mn}$ corresponding to $|\bar{\bf p}|_m = 0$ become singular. 
This is not a significant problem, since also the kernel $\widetilde{\cal K}_{\bar{p}, \bar{k}}$ vanishes for $|\bar{\bf p}| = 0$ or $|\bar{\bf k}| = 0$.
Therefore the singular basis functions can be neglected in the spectral decomposition of the kernel and in the computation of the collision integral.

One of the non-trivial features of the kernel, which is hard to reproduce, is the presence of a peak located at $|\bar{\bf p}| = |\bar{\bf k}|$ and $\cos \theta_{\bar p} = \cos \theta_{\bar k}$, whose height diverges for $\cos \theta_{\bar p} = \cos \theta_{\bar k} = \pm 1$. Such a peak originates from the forward scattering of the incoming particles.
We regularized such behaviour by setting an upper cut on the value of the kernel in $\cos\theta_{\bar p, \bar k} = \pm 1$. 
We checked that this approximation improves the numerical stability of the computation and has a negligible impact on the final result.

In our implementation we used a linearly-increasing spacing (i.e.~a quadratic distribution) for the points along the $|\bar{\bf p}|$ direction and a uniform spacing for the points along $\cos \theta_{\bar{p}}$. The non-uniform spacing in the momentum direction is motivated by the fact that it allows us to obtain a finer spacing at small momentum, where the kernel structure shows more complex features. This choice also helps in obtaining a more uniform reconstruction of the peak structure, whose width scales as $|\bar{\bf p}|$.

\begin{figure}
\centering
\includegraphics[width=.52\textwidth]{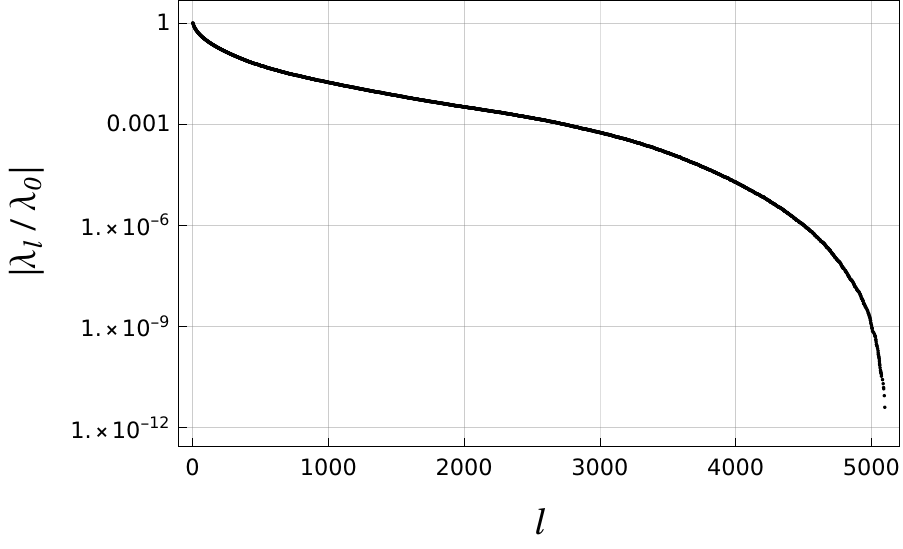}
\caption{Relative size of the eigenvalues of the kernel matrix ${\cal U}$ with $M = 100$ and $N = 51$. A large fraction of them (the first $\sim 3500$ eigenvalues) has a size $\gtrsim 10^{-4} |\lambda_0|$.
}\label{fig:eig_distr}
\end{figure}

For our numerical analysis we chose a grid with $M = 100$ (with the restriction $|\bar{\bf p}|/T \leq 20$) and $N = 51$ points, and we excluded the points with $|\bar p| = 0$. In this way the ${\cal U}$ matrix has dimension $5100$. 
We show in Fig.~\ref{fig:eig_distr} the relative size of the kernel eigenvalues in decreasing absolute value. The plot shows that, after a relatively fast decrease, the size of the eigenvalues tends to decrease slowly, so that a large fraction of them has a size $\gtrsim 10^{-4} |\lambda_0|$. As a consequence, in order to reconstruct the kernel with an overall accuracy of order $1-2\%$, almost all eigenvectors must be taken into account. Including in the sum only the largest $4000$ eigenvalues, the typical reconstruction error is $2-5\%$, while with $3000$ eigenvalues it grows to $5-10\%$.

We mention that significantly larger relative reconstruction errors are present in regions where the kernel is highly suppressed (such as for configurations in which $p$ is small while $k$ is large, or vice versa), or, especially for large momenta, around the peak. These regions, however have a limited impact on the computation of the collision integral, as confirmed by the numerical analysis reported in section~\ref{sec:analysis}.

By adopting the basis of eigenfunctions $\{\psi_l\}$ to decompose the perturbation, the computational time of the brackets is highly reduced. Compared to the timing performance of the method presented in ref.~\cite{DeCurtis:2022hlx} the decomposition method outlined in this section is two orders of magnitude faster. Such numerical improvement allows one to solve the Boltzmann equation in less than one hour on a desktop computer.

\section{Hydrodynamic equations for the background} \label{sec:hydro}

 In this section we focus on the characterization of the plasma that interacts with the DW. As we stated before, we consider the plasma to be a mixture of two different fluids: the top quark fluid, and the massless background, which contains all the species with negligible couplings to the DW. The out-of-equilibrium deviations of the latter are suppressed by the large number of degrees of freedom and can be neglected in a first approximation. In ref.~\cite{DeCurtis:2022hlx} we regarded the background as a thermal bath with temperature and velocity that do not depend on the position. In the present work we relax this approximation and we consider the massless species to be only in local equilibrium. In such a case, the background plasma is described by equilibrium distribution functions with position-dependent temperature and velocity.

The simplest approach to determine the temperature and velocity profiles for the plasma
relies on a linearization procedure but it has the major drawback to break down for $v_w = c_s$, where $c_s$ is the speed of sound in the plasma~\cite{Moore:1995si,Dorsch:2021nje}. To fully take into account the non-linearities and thus to avoid the singularity at $c_s$, one can exploit a set of hydrodynamic equations~\cite{Laurent:2022jrs} that are obtained from the conservation of the energy-momentum tensor of the system. We will briefly summarize this approach in the following.

The energy-momentum tensor can be conveniently split into three components: the one due to the scalar fields $\phi_i$ participating in the PhT ($T_\phi^{\mu\nu}$), the local-equilibrium contributions of all the species in the plasma ($T_{pl}^{\mu\nu}$), and the out-of-equilibrium deviations of the massive species ($T_{out}^{\mu\nu}$), namely
	\begin{equation}
		T^{\mu\nu} = T_\phi^{\mu\nu} + T_{pl}^{\mu\nu}+ T_{out}^{\mu\nu} \,.
	\end{equation}
In the reference frame of the DW, the conservation of the energy-momentum tensor leads to the following equations
	\begin{equation}\label{eq:tensor_conservation}
		\begin{split}
			&T^{30} = w(\phi_i, T)\gamma_{p}^2v_{p}+T^{30}_{out}= c_1\,,\\
			&T^{33} = \frac{1}{2}(\partial_z\phi_i)^2 - V(\phi_i, T)+ w\gamma_{p}^2v_{p}^2+T^{33}_{out} = c_2\,,
		\end{split}
	\end{equation}
where $V(\phi_i, T)$ is the finite-temperature effective potential, while $w(\phi_i,T)$ is the enthalpy defined as
	\begin{equation}
		w(\phi_i,T)=T\f{\partial V(\phi_i,T)}{\partial T}\,.
	\end{equation}
The two constants $c_1$ and $c_2$ can be determined from the boundary values of $v_p$ and $T$ far in front or far behind the DW.
 
Equations~(\ref{eq:tensor_conservation}) can be recast in the following form	
	\begin{equation}
		\begin{split}
			&v_{p}=\frac{- w(\phi_i,T) +\sqrt{4(c_1-T^{30}_{out})^2 + w^2}}{2(c_1-T^{30}_{out})}\,,\\
			\rule{0pt}{1.75em}&\frac{1}{2}(\partial_z\phi_i)^2 - V(\phi_i, T) -\frac{1}{2}w(\phi_i,T) + \frac{1}{2}\sqrt{4(c_1-T^{30}_{out})^2+w(\phi_i, T)^2} - (c_2-T^{33}_{out}) = 0\,,
		\end{split}
		\label{eq:bg_profiles}
	\end{equation}
and their solutions yield the velocity and temperature profiles. These equations can be straightforwardly solved with a simple root-finding numerical algorithm.

Finally, the boundary values of the temperature, the plasma velocity and the field VEVs far in front and far behind the DW, which we denote with a $+$ and $-$ subscript respectively, can be determined as described in ref.~\cite{Espinosa:2010hh}.
The fields VEVs can be easily found by minimizing the potential, namely,
	\begin{equation}
		\f{\partial V(\phi_{i\pm},T_{\pm})}{\partial\phi_i}=0 \,,
	\end{equation}
while the computation of $T_{\pm}$ and $v_{p\pm}$, that depend on the DW velocity $v_w$, is much more involved and is briefly outlined in the appendix (we refer to refs.~\cite{Espinosa:2010hh,Laurent:2022jrs} for further details).

\section{Numerical analysis}\label{sec:analysis}

To validate the method described in section~\ref{sec:decomposition} we compared it with the one we developed in our previous work~\cite{DeCurtis:2022hlx}.
In particular we compared the relevant quantities that enter in the computation of the DW terminal speed, namely the out-of-equilibrium corrections to the stress-energy tensor $T^{30}_{out}$, $T^{33}_{out}$,
and the friction $F(z)$.

We choose as a benchmark scenario the $Z_2$-symmetric singlet extension of the SM. This choice is motivated by the fact that the presence of a new scalar field $s$ affects the thermal history of the Universe and can give rise to a first-order EWPhT. The extra scalar, singlet under the SM gauge group, also allows for a two-step PhT, in which case the EW symmetry breaking is preceded by a $Z_2$-symmetry breaking in the extra sector (see for instance refs.~\cite{DeCurtis:2019rxl,Friedlander:2020tnq} and references therein). 

The tree-level potential of the model is
\begin{equation}
V_{tree}(h,s) = \frac{\lambda_h}{4} \left(h^2- v_0^2\right)^2+\frac{\lambda_s}{4}\left(s^2-w_0^2\right)^2 + \frac{\lambda_{hs}}{4} h^2s^2\,,
\end{equation}
where $v_0$ is the Higgs VEV at the EW minimum, $\lambda_h$ is the Higgs self coupling while $\lambda_s,$ $w_0$ and $\lambda_{hs}$ describe the singlet self-coupling, its VEV when the EW symmetry is exact and the portal coupling with the Higgs, respectively. The parameter $w_0$ can be traded for the physical mass of the singlet using the relation
\begin{equation}
   m_s^2 = -\lambda_s w_0^2 + \frac{1}{2}\lambda_{hs} v_0^2 \,.
\end{equation}
The finite-temperature effective potential of the model takes into account, besides the tree-level term, the one-loop corrections at zero temperature $V_1$, the counterterms that regularize the UV divergences $V_{CT}$ (we adopted the ${\overline {\rm MS}}$ renormalization scheme) and the thermal corrections $V_T$ (see ref.~\cite{Quiros:1999jp} for the details):
\begin{equation}
    V(h,s,T) = V_{tree}(h,s) + V_1(h,s) + V_{CT}(h,s) + V_T(h,s,T) \,.
\end{equation}
 Once the model parameters are chosen, the temperature and velocity profiles of the background plasma are computed by using the conservation laws of the stress-energy tensor. In addition, by solving the coupled system of the Boltzmann equation and the equations of motion of the scalar fields, one can determine the terminal speed of the DW.

\begin{figure}
    \centering
    {\includegraphics[width=.32\textwidth]{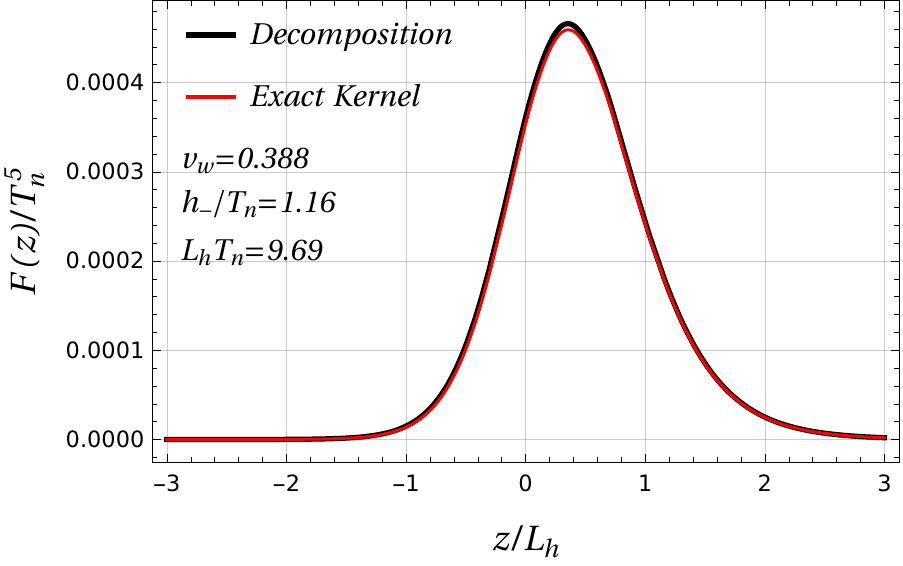}}
    \hfill
    {\includegraphics[width=.31\textwidth]{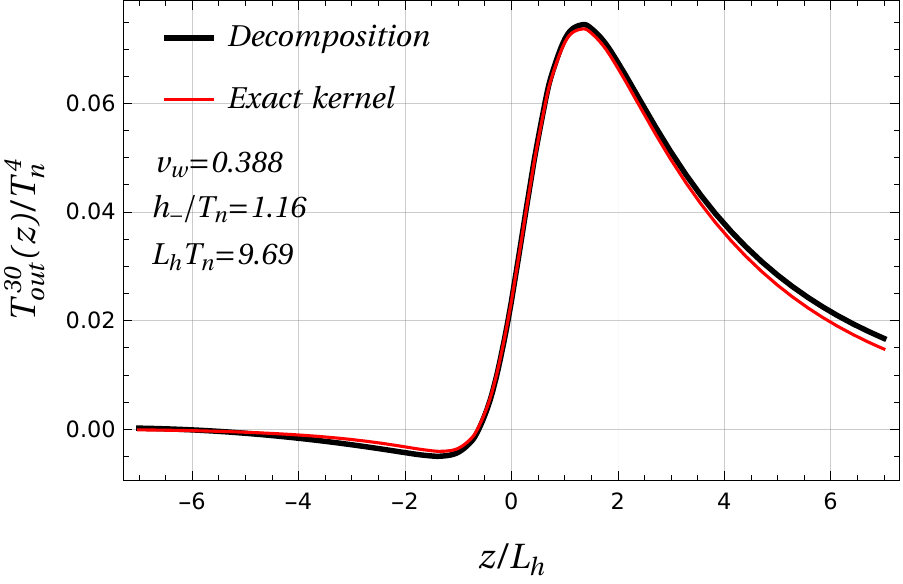}}
    \hfill
    {\includegraphics[width=.31\textwidth]{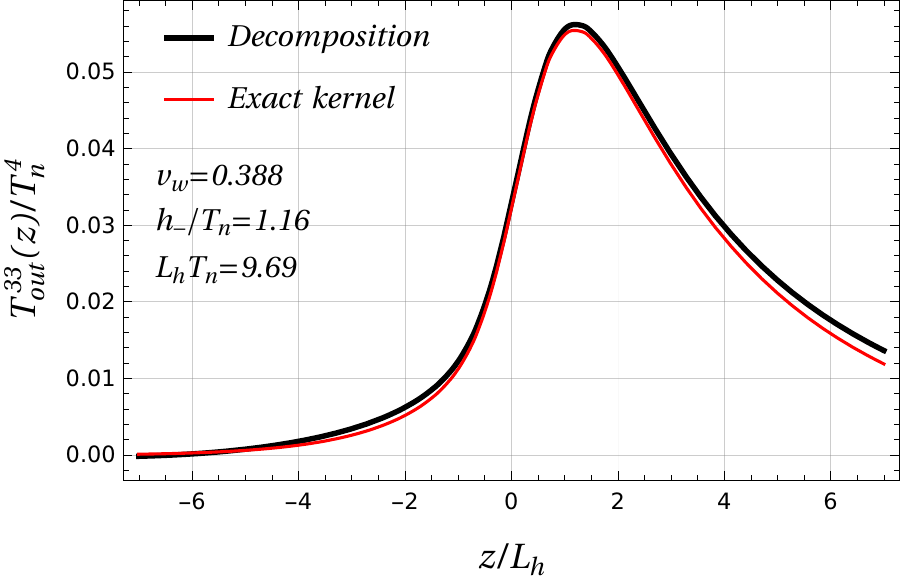}}
    \vspace{.5em}
    {\includegraphics[width=.32\textwidth]{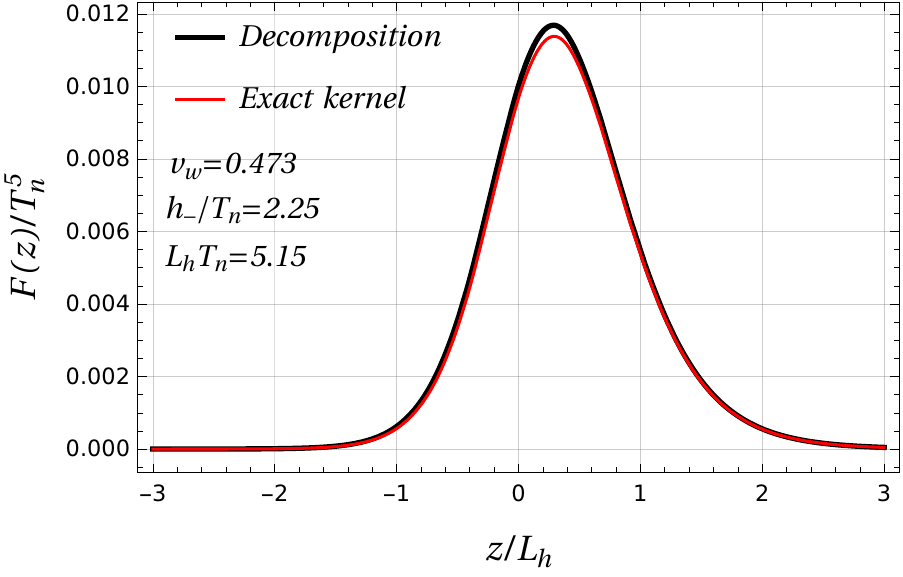}}
    \hfill
    {\includegraphics[width=.31\textwidth]{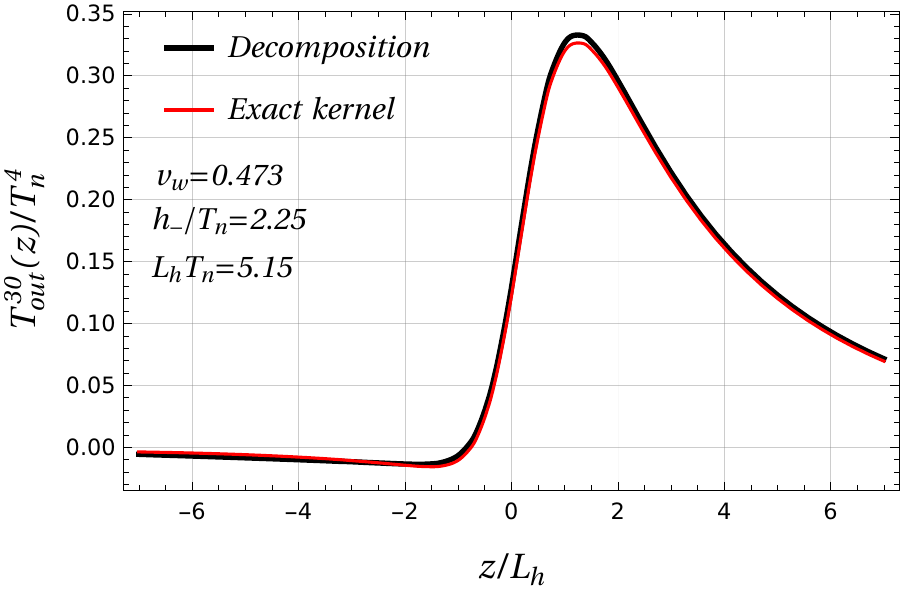}}
    \hfill
    {\includegraphics[width=.31\textwidth]{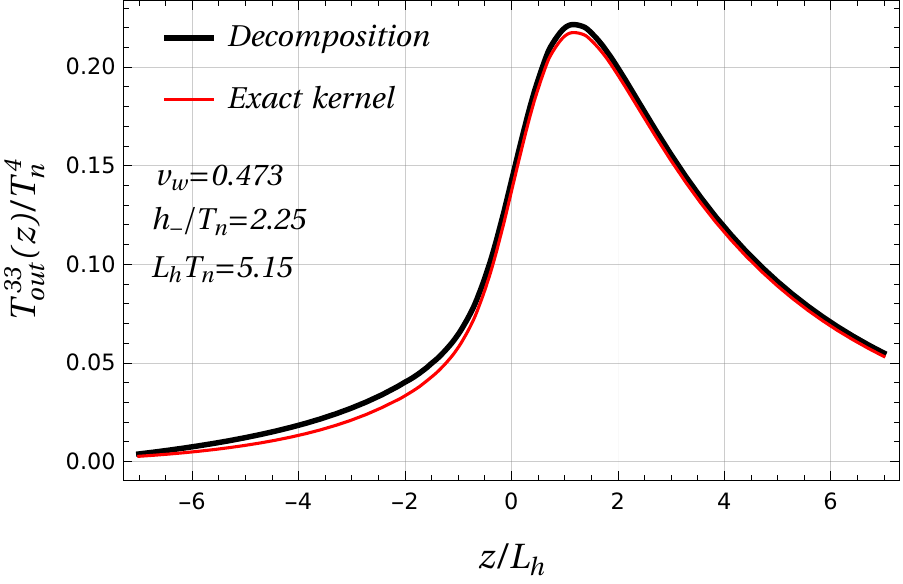}}
    \caption{Comparison of the friction (left panel), $T^{30}_{out}$ (central panel) and $T^{33}_{out}$ (right panel) as functions of $z/L_h$, computed using the procedure presented in ref.~\cite{DeCurtis:2022hlx} (red solid line) and the decomposition method (black solid line), for the benchmark point BP1 (upper row) and BP2 (lower row). The parameters defining the two benchmarks are reported in Tab.~\ref{tab:parameters_results}.}\label{fig:old_vs_new_compare}
\end{figure}

 In Fig.~\ref{fig:old_vs_new_compare} we plot $F(z)$, $T_{out}^{33}(z)$, $T_{out}^{30}(z)$ as functions of the ratio $z/L_h$, for two benchmark points
reported in Tab.~\ref{tab:parameters_results} and characterized by $v_w = 0.388$, $L_h\, T_n = 9.69$, $h_-/T_n = 1.16$ for the benchmark BP1 and by $v_w = 0.473$, $L_h\, T_n = 5.15$, $h_-/T_n = 2.25$ for BP2. As we will see in the following section, such values correspond to the terminal ones obtained for a scalar potential with parameters $m_s = 103.8 \, \textrm{GeV}$, $\lambda_{hs} = 0.72$ and $\lambda_s = 1$ for BP1 and $m_s = 80.0 \, \textrm{GeV}$, $\lambda_{hs} = 0.76$ and $\lambda_s = 1$ for BP2.
The plots clearly show that the new method correctly reproduces
the friction and the out-of-equilibrium corrections to the stress-energy tensor in the whole range of $z$. Although at the qualitative level an excellent agreement is found, some small quantitative differences are present. In particular, differences of order $1\%$
are present in the region corresponding to the peak ($z \sim 1$), while order $10\%$ discrepancies can be present in the tails, both inside and outside the bubble. The impact of the tails in the determination of the total friction and of the DW dynamics is however very limited, since their contribution is highly suppressed, as can be seen from the first panel in the two rows of Fig.~\ref{fig:old_vs_new_compare}.\footnote{For some choices of the model parameters, the convergence of the solution in the region $p_z<0$ and $z\ll0$ may present some numerical instabilities due to our approximation of the kernel. In this region, however, the perturbations are highly suppressed and give only a marginal contribution to the friction. Moreover possible numerical instability only appear after a number of iterative steps much larger than the ones needed to reach convergence in the relevant part of the perturbations, so they can be easily kept under control in the determination of the DW velocity.
}

These results thus indicate that the proposed method is robust and can be safely used to reliably study the DW dynamics.

\subsection{Solution to the scalar field equations of motion}\label{subsec:eom_scalars}

The DW terminal speed is the result of a balance between the internal pressure of the wall and the friction. 
To determine it, one has to solve the equation of motion of the two scalar fields together with the Boltzmann equation that determines the out-of-equilibrium perturbations and, through eq.~(\ref{eq:friction}), the friction.

In the wall reference frame the equations of motion of the Higgs and scalar singlet fields are
\begin{equation}\label{eq:eom_scalars}
\begin{split}
    E_h &\equiv -\partial_z^2 h + \frac{\partial V(h,s,T)}{\partial h} + F(z)/h' = 0 \,, \\
    \rule{0pt}{1.75em}E_s &\equiv -\partial_z^2 s + \frac{\partial V(h,s,T)}{\partial s} = 0  \,.
\end{split}
\end{equation}
The solution to the above system of equations yields the exact profiles of the Higgs and singlet fields. For most applications, however, it is not necessary to solve for the exact field profiles. A good approximate solution
is given by a $\tanh$ ansatz, namely
\begin{equation}\label{eq:fieldansatz}
\begin{split}
    h(z) &= \frac{h_-}{2}\left(1+\tanh\left(\f{z}{L_h}\right)\right)\\
    \rule{0pt}{1.75em}s(z) &= \f{s_+}{2}\left(1-\tanh\left(\f{z}{L_s}+\delta_s\right)\right)
\end{split}
\end{equation}
where $L_h$ and $L_s$ are the wall thicknesses of the Higgs and the scalar singlet, respectively,
while the parameter $\delta_s$ describes the displacement between the Higgs and the singlet field walls.
The values of the VEVs of the Higgs field inside the DW, $h_-$, and the singlet in front of the DW, $s_+$, are computed by minimizing the finite-temperature effective potential, namely
\begin{equation}
    \f{\partial V(h_-,0,T_-)}{\partial h} = 0\,, \qquad \quad
    \f{\partial V(0,s_+,T_+)}{\partial s} = 0\,.
\end{equation}
Clearly, it is possible to consider more general field profiles,
but the ansatz in eq.~(\ref{eq:fieldansatz}) is sufficient to reproduce the relevant features of the DW dynamics~\cite{Laurent:2022jrs, Friedlander:2020tnq}.

The standard strategy to determine $v_w,\, L_h,\, L_s$ and $\delta_s$ is to compute momenta of $E_h$ and $E_s$ and then seek for the roots of the corresponding equations. A convenient choice of the momenta is~\cite{Laurent:2022jrs, Friedlander:2020tnq} 
\begin{equation}\label{eq:param_equations}
\begin{split}
    P_h & = \int dz\, E_h h' = 0\,, \qquad \quad G_h = \int dz\, E_h (2 h/h_- - 1)h' = 0\,, \\
    \rule{0pt}{1.75em}P_s & = \int dz\, E_s s' = 0\,, \qquad \quad G_s = \int dz E_s\, (2 s/s_+ -1)s' = 0\,.
\end{split}
\end{equation}
These equations have a clear physical interpretation. $P_{h,s}$ correspond to the total pressures on the two walls. Thus the sum $P_h + P_s$ is the total pressure that acts on the system. Both pressures must vanish in the steady state regime, otherwise the walls would accelerate. As confirmed by the numerical analysis, we expect the combination $P_h+P_s$ to mostly depend on $v_w$ and being less sensitive to the remaining parameters. The difference of pressures $P_h-P_s$, instead, depends mainly on the displacement $\delta_s$. Finally $G_{h,s}$ take into account the presence of pressure gradients which may stretch or compress the walls. Thus the requirement $G_{h,s} = 0$ fixes the widths of the two walls.

In order to close the system of equations~(\ref{eq:param_equations}), one should include the Boltzmann equation for the out-of-equilibrium distributions. However, the perturbations that determine the friction depend on the unknown parameters $v_w$ and $L_h$. To address this problem one can employ the iterative procedure outlined below:
\begin{enumerate}
\item Solve eq.~(\ref{eq:param_equations}) without the perturbations, namely at equilibrium, and compute the four parameters $v_w$, $L_h$, $L_s$ and $\delta_s$.
\item Use the four parameters to determine the out-of-equilibrium perturbations by solving the Boltzmann equation.
\item Insert the perturbations in eq.~(\ref{eq:eom_scalars}) and recompute the parameters from eq.~(\ref{eq:param_equations}).
\item Iterate the procedure from point 2 until the convergence of the four parameters is reached.
\end{enumerate}

\begin{table}[]
    \centering
    \begin{tabular}{c|c|c|c||c|c|c|c}
     & $m_s\,$(GeV) & $\lambda_{hs}$ & $\lambda_s$ & $T_n\,$(GeV) & $T_c\,$(GeV) & $T_+\,$(GeV) & $T_-\,$(GeV)   \\
     \hline
     \rule{0pt}{1.1em}BP1 & 103.8 & 0.72 & 1 & 129.9 & 132.5 & 130.3 & 129.9 \\ 
     \rule{0pt}{1.em}BP2 & 80.0 & 0.76 & 1 & 95.5 & 102.8 & 97.5 & 95.5
    \end{tabular}
    \\
    \vspace{0.5cm}
    \begin{tabular}{c|c|c|c|c}
     & $v_w$ & $\delta_s$ & $L_h T_n$ & $L_s T_n$ \\
     \hline
     \rule{0pt}{1.1em}BP1 & 0.39\ \ (0.57) & 0.79\ \ (0.75) & 9.7\ \ (8.1) & 7.7\ \ (6.7)\\ 
     \rule{0pt}{1.em}BP2 & 0.47\ \ (0.61) & 0.81\ \ (0.81) & 5.2\ \ (4.7) & 4.3\ \ (4.1)
    \end{tabular}
    \caption{Critical and nucleation temperatures, temperatures in front and behind the DW and terminal values of the parameters $v_w$, $\delta_s$, $L_h$, $L_s$ for two benchmark points.
    The numbers in parentheses correspond to the results obtained neglecting the out-of-equilibrium perturbations.}
    \label{tab:parameters_results}
\end{table}

To numerically solve eq.~(\ref{eq:param_equations}) we implemented the Newton algorithm, while to solve the Boltzmann equation we followed the iterative procedure described in ref.~\cite{DeCurtis:2022hlx} with the bracket term $\langle \delta f\rangle$ computed following the strategy explained in section~\ref{sec:decomposition}. We report in Tab.~\ref{tab:parameters_results} our results for two benchmark points. In the table we show the values of the four parameters with and without the contributions of the out-of-equilibrium perturbations. 
The bubble speed is the parameter more strongly affected by the out-of-equilibrium contributions, followed by the width of the Higgs wall.
For the first benchmark model with $m_s = 103.8$ GeV, a difference of $\sim 30\%$ and of $\sim 20\%$ is present for the speed $v_w$ and the wall width $L_h$, respectively, with respect to the same values computed in local equilibrium. The width $L_s$ and the displacement $\delta_s$ show, instead, a difference of $\sim 15\%$ and a milder one of $\sim 5\%$ respectively. 
The four parameters in the second benchmark model with $m_s = 80$ GeV still present important differences with respect to the only-equilibrium case, but the impact of the out-of-equilibrium contributions is less severe. As for the previous benchmark model, the perturbations mostly impact on the speed on which they induce a change of $\sim 20\%$. 
The offset $\delta_s$, instead, is almost unaffected by the inclusion of the out-of-equilibrium corrections.

To understand why the out-of-equilibrium perturbations have a different impact on the two benchmark models, it is useful to study the total pressure acting on the system. This will also clarify why the terminal speed $v_w$ is the parameter more strongly affected by the out-of-equilibrium corrections. Using eq.~(\ref{eq:param_equations}) the total pressure acting on the system can be expressed as
\begin{equation}\label{eq:total_pressure}
    P_h + P_s =\Delta V + \int dz \,\f{T'}{T}w(T) + \int dz\, F(z) = 0
\end{equation}
where $\Delta V$ is the potential energy difference between the true and false vacuum. The size of this difference increases with the amount of supercooling in the PhT. For small supercooling, the friction is comparable to the potential difference and the out-of-equilibrium corrections have an important impact on the DW dynamics. As we notice from Tab.~\ref{tab:parameters_results}, the model where the out-of-equilibrium corrections have a bigger impact indeed corresponds to the one with less supercooling.

\begin{figure}
    \centering
    \includegraphics[width=0.32\textwidth]{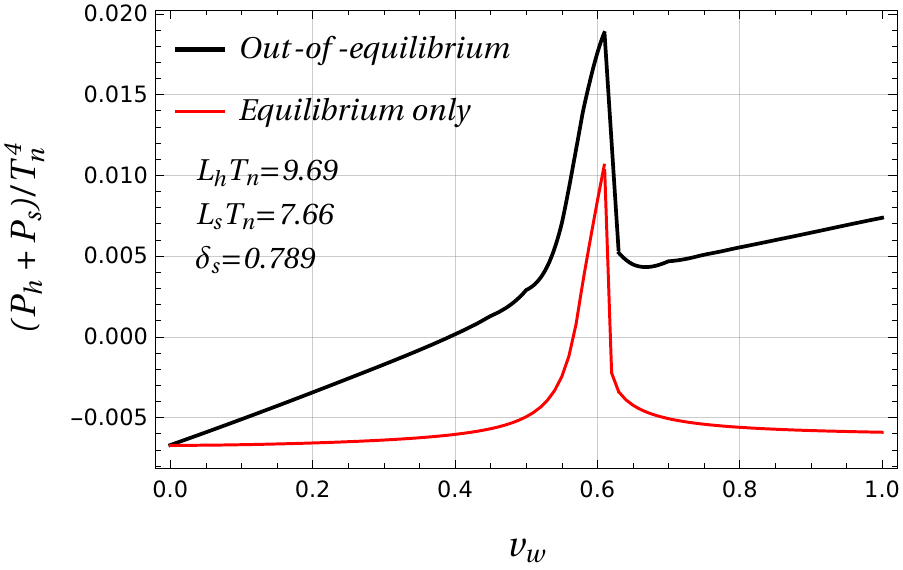}
    \hfill
    \includegraphics[width=0.32\textwidth]{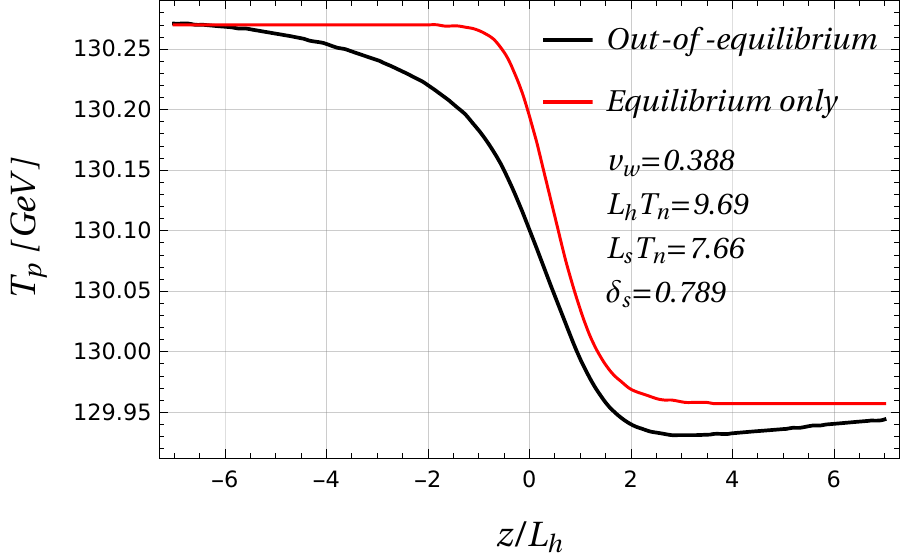}
    \hfill
    \includegraphics[width=0.32\textwidth]{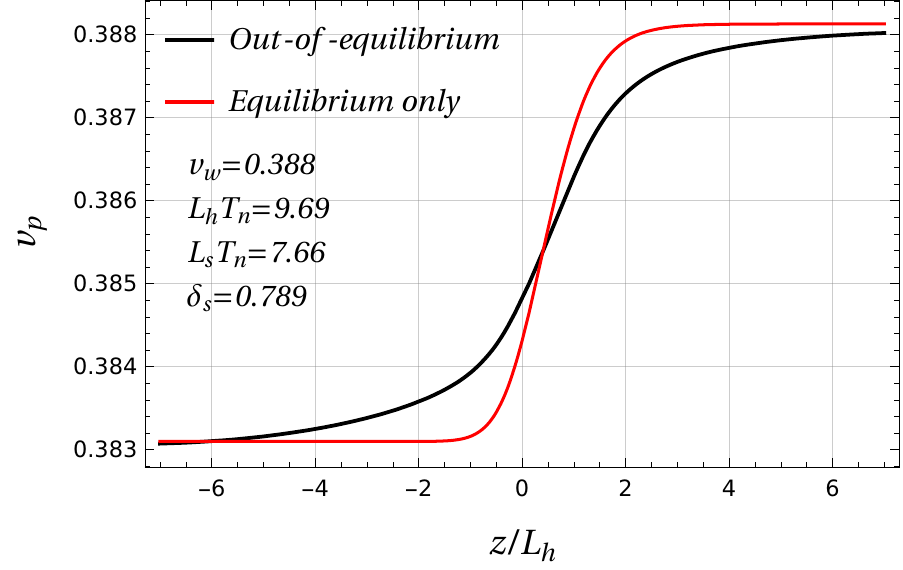}\\
    \vspace{.5em}    
    \includegraphics[width=0.32\textwidth]{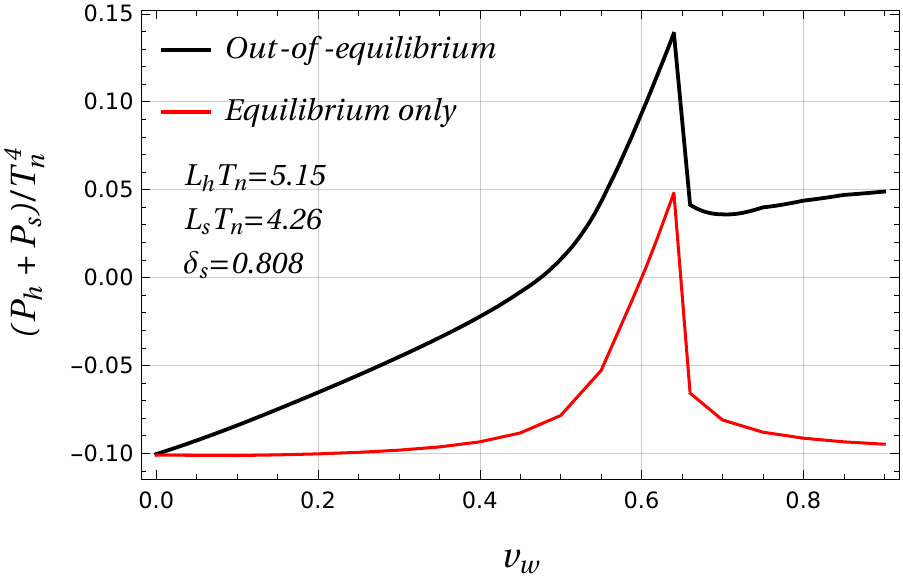}
    \hfill
    \includegraphics[width=0.31\textwidth]{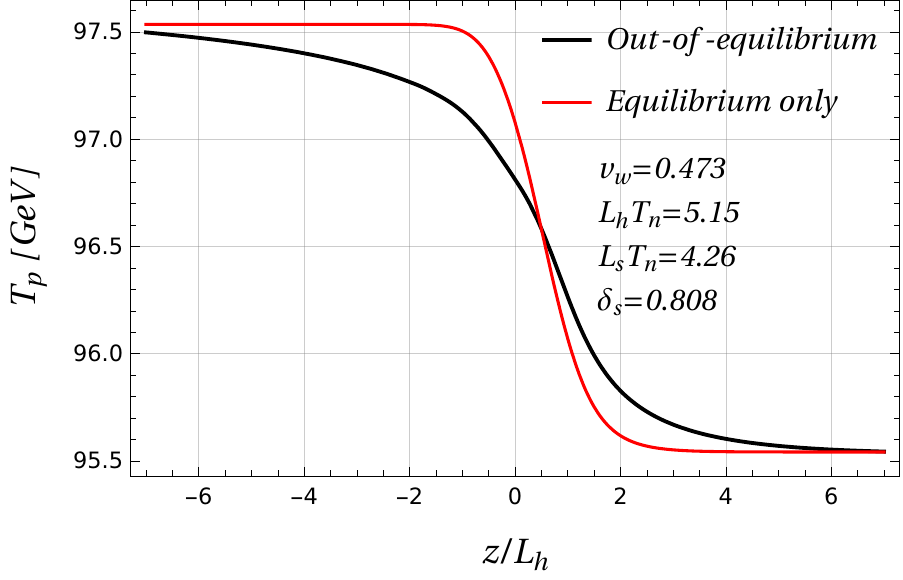}
    \hfill
    \includegraphics[width=0.32\textwidth]{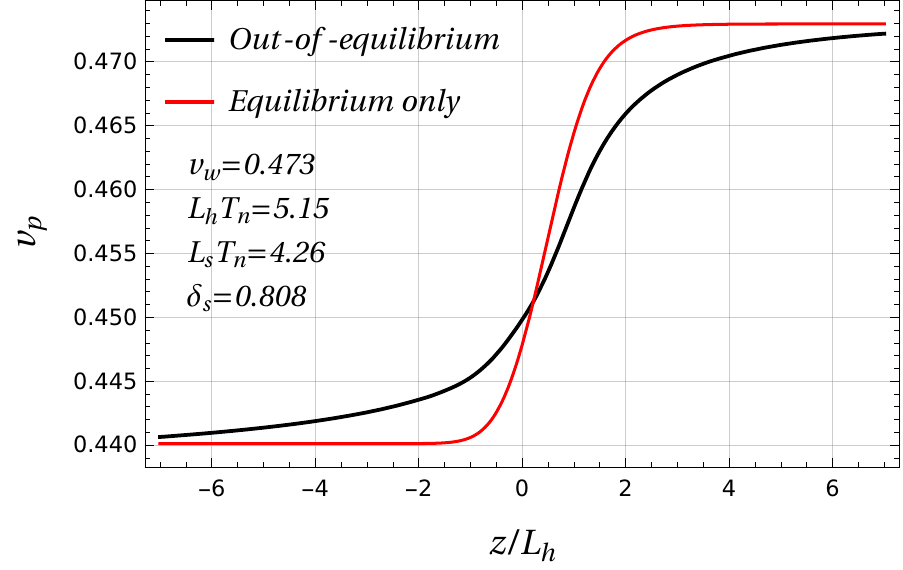}
    \caption{Total pressure as a function of the bubble speed, temperature and velocity profiles as functions of $z/L_h$ for the two benchmark models reported in Tab.~\ref{tab:parameters_results} (BP1 on the upper row, BP2 on the lower row). The red solid lines are obtained by neglecting the out-of-equilibrium perturbations, while the black solid lines correspond to the complete computation. The peak in the pressure is located at the Jouguet velocity.}
    \label{fig:pressure_and_profiles}
\end{figure}

We plot on the left panel of Fig.~\ref{fig:pressure_and_profiles} the total pressure as a function of the wall speed, with (black line) and without (red line) the inclusion of the out-of-equilibrium distributions. In both curves we can observe the presence of a peak corresponding to the Jouguet velocity. The peak originates from hydrodynamic effects that heat up the plasma, thus generating a pressure barrier that slows down the DW motion in models with small supercooling \cite{Konstandin:2010dm}. This effect is described in eq.~(\ref{eq:total_pressure}) by the term proportional to the enthalpy and its impact grows with the difference $T_+ - T_-$, so that it tends to be larger for hybrid walls. 

At small velocities, instead, the pressure at equilibrium becomes constant. In this case the temperature difference across the wall becomes negligible and the value of the total pressure settles to the potential energy difference between the false and true vacua at the nucleation temperature.

 The out-of-equilibrium perturbations provide a correction that grows linearly with the velocity. This behaviour is consistent with the one found in ref.~\cite{DeCurtis:2022hlx}, where a linear dependence of the total friction was observed. From the above discussion we conclude that the perturbations have an important impact on the wall terminal speed and, therefore, an accurate modeling of the out-of-equilibrium perturbations is necessary to get a proper description of the PhT dynamics.

In the central and right panel of Fig.~\ref{fig:pressure_and_profiles} we plot the temperature and velocity profiles, respectively, for the terminal values of $v_w,\,L_h,\,L_s,\,\delta_s$ of the two benchmark models reported in Tab.~\ref{tab:parameters_results}. The plots clearly show that the out-of-equilibrium perturbations impact on the shape of the profiles, mostly in the region close to the DW. Among all the three terms in eq.~(\ref{eq:total_pressure}), the modified shapes of the profiles have the largest effect on the term involving the temperature derivative.

We now give an estimation of the impact of the temperature and velocity profiles on the Boltzmann equation. They contribute to the source term through their derivatives and, thus, we expect their effect to be proportional to the relative difference between the temperature and velocity across the DW. This expectation is confirmed by our numerical analysis. We computed the integral of the friction for three benchmark velocities ($v_w=0.4, 0.6, 0.9$) and we compared with the result obtained by considering a constant temperature and plasma velocity across the system. Our analysis showed that corrections of few $\%$ are present for $v_w = 0.4$ and $v_w = 0.9$, while we found larger corrections ($\sim 10\%$) for $v_w = 0.6$. Such result agrees with the expected behavior since the largest correction is found for velocities close to the Jouguet velocity, where the differences between the temperature and plasma velocity across the wall are larger. From these results we can conclude that, despite the profiles have to be included to compute the equilibrium part of the total pressure acting on the system, they may be typically neglected in the computation of the out-of-equilibrium friction.

Finally we comment on the comparison of our results with the previous literature. We found a fair agreement with the results given in ref.~\cite{Friedlander:2020tnq}, in which the DW terminal speed for a similar potential is computed
by solving the Boltzmann equation within the fluid approximation. This approach, as we showed in ref.~\cite{DeCurtis:2022hlx}, tends to overestimate the value of the friction for subsonic walls by $\sim 10\% - 20\%$. The results reported in Tab.~\ref{tab:parameters_results} show differences with respect to the values in ref.~\cite{Friedlander:2020tnq} that are compatible up to these effects.

On the other hand, we found much larger differences with respect to the results in ref.~\cite{Laurent:2022jrs}, in which the out-of-equilibrium perturbations are found to provide only a small correction to the DW terminal velocity. To investigate the source of discrepancy, we explicitly implemented the procedure proposed in ref.~\cite{Laurent:2022jrs}, decomposing the perturbations in terms of Chebyschev polynomials. With this approach we recovered the results obtained with our method up to $\sim 20\%$ differences, which are most probably due to the different size of the grids used to discretize the perturbations. Other sources of differences could be the choice of the scalar potential (and of its renormalization procedure) and of the model parameters, which could lead to a different amount of supercooling.

 \section{Conclusions}\label{sec:concl}

In this work we developed a new algorithm for the computation of the collision integrals in the Boltzmann equations that determine the distribution functions of plasma species in the presence of a travelling DW. This physical setup is relevant for a quantitative description of first-order PhTs in the early Universe and, in particular, for the computation of cosmic relics, such as the stochastic background of gravitational waves and the matter-antimatter asymmetry.

The method presented here exploits a spectral decomposition of the collision integral, reinterpreting it as a suitable Hermitian integral operator.
This new approach significantly improves the computational efficiency of our previous algorithm~\cite{DeCurtis:2022hlx} by effectively reducing a nine-dimensional integration into a much faster matrix multiplication. On top of that, since the eigenfunctions need to be computed only once, they can be reused in all the scanning procedure of the parameter space of an entire model. 

To validate our new spectral method we compared it with the approach we developed in our previous work~\cite{DeCurtis:2022hlx}. We found excellent agreement for the value of the friction and for the determination of the out-of-equilibrium perturbations in the plasma. Small deviations, of order $1\%$ in the most relevant regions (see Fig.~\ref{fig:old_vs_new_compare}), could be systematically reduced by increasing the number of points in the grid used for the discretization of the collision operator.

We further complemented our analysis by including the contribution of the background species obtained by solving the hydrodynamic equations for their temperature and velocity profiles. 
Contrary to most of the previous approaches, following ref.~\cite{Laurent:2022jrs}, we did not linearize in the former quantities, thus avoiding any singular behavior at the speed of sound. 

As an explicit application of our formalism, we applied it to the study of a first-order EWPhT in the $Z_2$-symmetric singlet extension of the SM. We inspected two benchmark points, calculating the relevant features of the wall dynamics, namely the wall speed, its width and the displacement between the two field profiles.
In order to asses the impact of the out-of-equilibrium distributions we also reevaluated the same parameters without the friction. The main results are shown in Fig.~\ref{fig:pressure_and_profiles} and in Tab.~\ref{tab:parameters_results}, and clearly highlight the crucial impact of the out-of-equilibrium dynamics, which introduces large corrections, mainly to the speed of the bubble wall. 

For the purpose of presenting the new computational strategy, we considered a simplified setup with the top quark being the only species out-of-equilibrium. Clearly, the inclusion of EW gauge bosons, or any other massive state, is straightforward once the kernels for the corresponding processes are computed. We leave this aspect for a future work.
	
Finally, we remark that the decomposition method of the collision integrals presented in this work increases the feasibility of a scan of the parameter space of an entire model that otherwise would be extremely challenging due to the amount of required computational resources.
As we mentioned before, the huge timing performance that we achieved allows to fully solve the Boltzmann equation efficiently and reliably. We plan to analyze some of the most relevant BSM scenarios providing signals of gravitational waves from cosmological first-order PhT in a following work.

\section*{Acknowledgments}
        We thank B. Laurent for useful discussions. This project was supported in part by the MIUR under contract 2017FMJFMW (PRIN2017). A.G.M. has been supported by the Secretariat for Universities and Research of the Ministry of Business and Knowledge of the Government of Catalonia and the European Social Fund.

\appendix
\section{Boundary conditions for the hydrodynamic equations} \label{sec:appendix}

In this appendix we briefly review the computation of the boundary values of the plasma velocity $v_{p\pm}$ and temperature $T_{\pm}$ (see refs.~\cite{Espinosa:2010hh,Laurent:2022jrs} for more details).

The conservation of the energy-momentum tensor across the DW implies, 
	\begin{equation}\label{eq:vpvm}
		v_{p+}v_{p-} =\frac{1-(1-3\alpha_+)r}{3-3(1+\alpha_+)r}\qquad \frac{v_{p+}}{v_{p-}}=\frac{3+(1-3\alpha_+)r}{1+3(1+\alpha_+)r},	
	\end{equation}
where $v_{p+}$ and $v_{p-}$ are computed in the wall reference frame and 
\begin{equation}
\alpha_+ = \frac{\epsilon_+-\epsilon_-}{a_+T_+^4} \,, \qquad 
r = \frac{a_+T_+^4}{a_-T_-^4} \,,
\end{equation}
with 
\begin{equation}
a_{\pm} = -\frac{3}{4T_\pm^3}\frac{\partial V}{\partial T}\Bigg|_{\phi_i=\phi_i^\pm,T=T^\pm} \,, \qquad
		\epsilon_{\pm} = \left(-\frac{T_\pm}{4}\frac{\partial V}{\partial T}+V\right)\Bigg|_{\phi_i=\phi_i^\pm,T=T^\pm} \,.
\end{equation}
By using eq.~(\ref{eq:vpvm}) we can express two of the four boundary conditions as functions of the remaining two. The latter depend on the qualitative properties of the plasma which are, in turn, determined by the DW velocity $v_w$. It is possible to identify three different regimes, depending on the value of $v_w$, for the temperature and plasma velocity profiles, each one characterized by different boundary conditions: deflagration ($v_w < c_s^-$) where a shock wave precedes the DW, detonations ($v_w > v_J$) where a rarefaction wave trails the DW and hybrid ($c_s^-<v_w<v_J$), where both a shock and a rarefaction wave are present. 
The speed of sound inside the wall $c_s^-$ can be computed from the definition of $c_s$, namely,
	\begin{equation}
		c_s^2 = \f{\partial V/\partial T}{T\partial^2 V/\partial T^2} \,,
	\end{equation}
while $v_J$ identifies the model-dependent Jouguet velocity, at which $|v_{p-}| = c_s^-$ holds.
	
In the case of detonations, the wall hits an unperturbed plasma in front of it and one can trivially identify $T_+ = T_n$ and $|v_{p+}| = v_w$ (in the wall frame), thus immediately determining, with the use of eq.~(\ref{eq:vpvm}), the four boundary conditions. 

The determination of the boundary conditions in the deflagration and hybrid cases is much more involved and requires to evolve the temperature and velocity profiles between the DW and the shock wave. Assuming a spherical bubble and a thin wall, from the conservation of the energy-momentum tensor we find, in the reference frame of the bubble center, 
	\begin{equation}\label{eq:sw_eq}
		\begin{split}
		2\frac{v_p(\xi)}{\xi}&=\gamma(v_p(\xi))^2(1-v_p(\xi)\xi)\left(\frac{\mu(\xi,v_p(\xi))^2}{c_s^2}-1\right)\partial_\xi v_p(\xi)\\
		\rule{0pt}{1.5em}\partial_\xi T(\xi)&=T(\xi)\gamma(v_p(\xi))^2\mu(\xi,v_p(\xi))\partial_\xi v_p(\xi),
		\end{split}
	\end{equation}
	where $\xi$ is the velocity of a given wave profile, $v_p(\xi)$ is the plasma velocity in the frame of the bubble center, while $\mu(\xi,v_p(\xi))$ is the Lorentz-transformed plasma velocity (in the reference frame of the wave profile)
	\begin{equation}
		\mu(\xi,v_p(\xi))=\f{\xi-v_p(\xi)}{1-\xi v_p(\xi)}.
	\end{equation}

In the deflagration case the shock wave hits an unperturbed plasma in front of it, $T_+^{SW} = T_n$, and the plasma is at rest behind the DW, $v_{p-} = v_w$, with $T_+^{SW}$ being the temperature in front of the shock wave. Thus the two boundary conditions are specified at different points in the plasma and the determination of $T_{\pm}$ and $v_{p+}$ is non-trivial. For this purpose, we adopted a shooting method which is sketched below:
\begin{enumerate}
\item We make an ansatz for $T_-$ we determine $T_+$ and $v_{p+}$ using eq.~(\ref{eq:vpvm});
\item We use them as initial conditions to integrate eq.~(\ref{eq:sw_eq}), provided $v_{p+}$ is boosted in the reference frame of the bubble center ($ v(\xi = v_w) = \mu(v_w,v_{p+})$);
\item We stop integration just behind the shock wave at $\xi_\textrm{SW}$, which is given by the condition $v_{p-}^\textrm{SW} \xi_\textrm{SW} = c_s^+ = 1/\sqrt{3}$; 
\item The outcome of the solution of eq.~(\ref{eq:sw_eq}), namely the velocity and temperature of the plasma behind the shock wave $v_{p-}^\textrm{SW},\,T_-^\textrm{SW}$, are used to compute $v_{p+}^\textrm{SW},\,T_+^\textrm{SW}$ by exploiting again eq.~(\ref{eq:vpvm});
\item The procedure is iterated adjusting $T_-$ until the condition $T_+^\textrm{SW} = T_n$ is reached.
\end{enumerate}
To determine the boundary conditions for hybrid solutions we can adopt the same strategy, with the only difference that entropy conservation enforces $v_{p-} = c_s^-$.

\providecommand{\href}[2]{#2}\begingroup\raggedright\endgroup

\end{document}